\documentclass[11pt,twocolumn]{article}
\usepackage[utf8]{inputenc} 
\usepackage{graphicx}
\usepackage{times}
\usepackage{amsmath}
\usepackage{amssymb}
\newcommand{\Heff}{H_\mathrm{eff}}

\newcommand{\ensav}[1]{\langle #1\rangle}
\newcommand{\ket}[1]{|#1\rangle}
\newcommand{\tstar}{\tau^*}
\newcommand{\tpre}{\tau_\mathrm{pre}}

\newcommand{\tTC}{\tau_{\textnormal{\scriptsize \textrm{PDTC}}}}

\newcommand{\down}{\ket{\!\!\downarrow}}
\newcommand{\up}{\ket{\!\!\uparrow}}

\title{Observation of a prethermal discrete time crystal}

\author
{\hspace{-00mm}A. Kyprianidis\footnote{These authors contributed equally to the preparation of this manuscript.}\:\:$^{1\dagger}$, F. Machado\footnotemark[\value{footnote}]\:\:$^{2,3}$, W. Morong$^{1}$, P. Becker$^{1}$, K. S. Collins$^{1}$, D. V. Else$^{4}$, \\
\hspace{-00mm} L. Feng$^{1}$, P. W. Hess$^{5}$, C. Nayak$^{6,7}$, G. Pagano$^{8}$, N. Y. Yao$^{2,3}$, and C. Monroe$^{1}$
\\
\normalsize{\hspace{-00mm}$^{1}$Joint Quantum Institute, Dept. of Physics and Joint Center for Quantum Information} \\ \normalsize{and Computer Science, University of Maryland, College Park, MD 20742 USA}\\
\normalsize{\hspace{-00mm}$^{2}$Dept. of Physics, University of California, Berkeley, CA 94720 USA}\\
\normalsize{\hspace{-00mm}$^{3}$Materials Sciences Division, Lawrence Berkeley National Laboratory, Berkeley, CA 94720, USA}\\
\normalsize{\hspace{-00mm}$^{4}$Dept. of Physics, Massachusetts Institute of Technology, Cambridge, MA 02139 USA}\\
\normalsize{\hspace{-00mm}$^{5}$Dept. of Physics, Middlebury College, Middlebury, VT 05753 USA}\\
\normalsize{\hspace{-00mm}$^{6}$Microsoft Quantum, Station Q, Santa Barbara, CA 93106 USA}\\
\normalsize{\hspace{-00mm}$^{7}$Dept. of Physics, University of California, Santa Barbara, CA 93106 USA}\\
\normalsize{\hspace{-00mm}$^{8}$Dept. of Physics and Astronomy, Rice University, Houston, TX  77005 USA}\\

\\
\normalsize{$^\dagger$To whom correspondence should be addressed; E-mail:  akyprian@umd.edu.}
}

\date{}
\begin{document}

\maketitle 

\textbf{
The conventional framework for defining and understanding phases of matter requires thermodynamic equilibrium. 
Extensions to non-equilibrium systems have led to surprising insights into the nature of many-body thermalization and the discovery of novel phases of matter, often catalyzed by driving the system periodically. 
The inherent heating from such Floquet drives can be tempered by including strong disorder in the system, but this can also mask the generality of non-equilibrium phases.
In this work, we utilize a trapped-ion quantum simulator to observe signatures of a non-equilibrium driven phase without disorder: the prethermal discrete time crystal (PDTC).
Here, many-body heating is suppressed not by disorder-induced many-body localization, but instead via high-frequency driving, leading to an expansive time window where non-equilibrium phases can emerge. 
We observe a number of key features that distinguish the PDTC from its many-body-localized disordered counterpart, such as the drive-frequency control of its lifetime and the dependence of time-crystalline order on the energy density of the initial state. 
Floquet prethermalization is thus presented as a general strategy for creating, stabilizing and studying intrinsically out-of-equilibrium phases of matter.  
}


The periodic modulation of a system represents a versatile technique for controlling its behavior, enabling the emergence of phenomena ranging from parametric synchronization to dynamic stabilization~\cite{strogatz2018nonlinear,Landau1976Mechanics}. 
Periodic driving has become a staple in fields ranging from  nuclear magnetic resonance  spectroscopy to quantum information processing  \cite{Mansfield_1971,Vandersypen_Chuang_2005, Zhou_Choi_Choi_Landig_Douglas_Isoya_Jelezko_Onoda_Sumiya_Cappellaro_etal_2020, Choi_Zhou_Knowles_Landig_Choi_Lukin_2020}.
On a more fundamental level, such a periodic Floquet drive also causes the system to exhibit a discrete time-translational symmetry.
Remarkably, this symmetry can be spontaneously broken to form time-crystalline order and can also be utilized to protect novel Floquet topological phases~\cite{Oka_Kitamura_2019,Potter_2016, Nathan_2019, Else_Bauer_Nayak_2016, Khemani_Lazarides_Moessner_Sondhi_2016, Yao_Potter_Potirniche_Vishwanath_2017}. 

The realization of many-body Floquet phases of matter requires overcoming two crucial challenges.
First, the system must not absorb energy from the driving field. 
In the presence of a periodic drive, dynamics are not constrained by energy conservation, and Floquet heating causes a generic many-body system to approach infinite temperature, precluding the existence of any non-trivial order~\cite{DAlessio_Rigol_2014}.
Second, genuine late-time dynamics must be clearly differentiated from early-time transient behavior: a phase of matter can only be characterized after dynamical processes lead to steady state behavior.

The conventional strategy for addressing the first (heating) challenge is to utilize strong disorder to induce many-body localization (MBL), where the presence of an extensive set of conserved local quantities prevents Floquet heating~ \cite{Nandkishore_Huse_2015, Ponte_2015, Abanin_Altman_Bloch_Serbyn_2019}.
However, requiring many-body localization leads to difficulties, including stringent constraints on both the dimensionality and the range of interactions \cite{Yao_Laumann_Gopalakrishnan_Knap_Muller_Demler_Lukin_2014, DeRoeck_Huveneers_2017}.
Moreover, the presence of strong disorder further slows down equilibration, making it even more difficult to overcome the second (timescale) challenge and distinguish between early- and late-time dynamics. 


\begin{figure*}
\centering
\includegraphics[width=\textwidth]{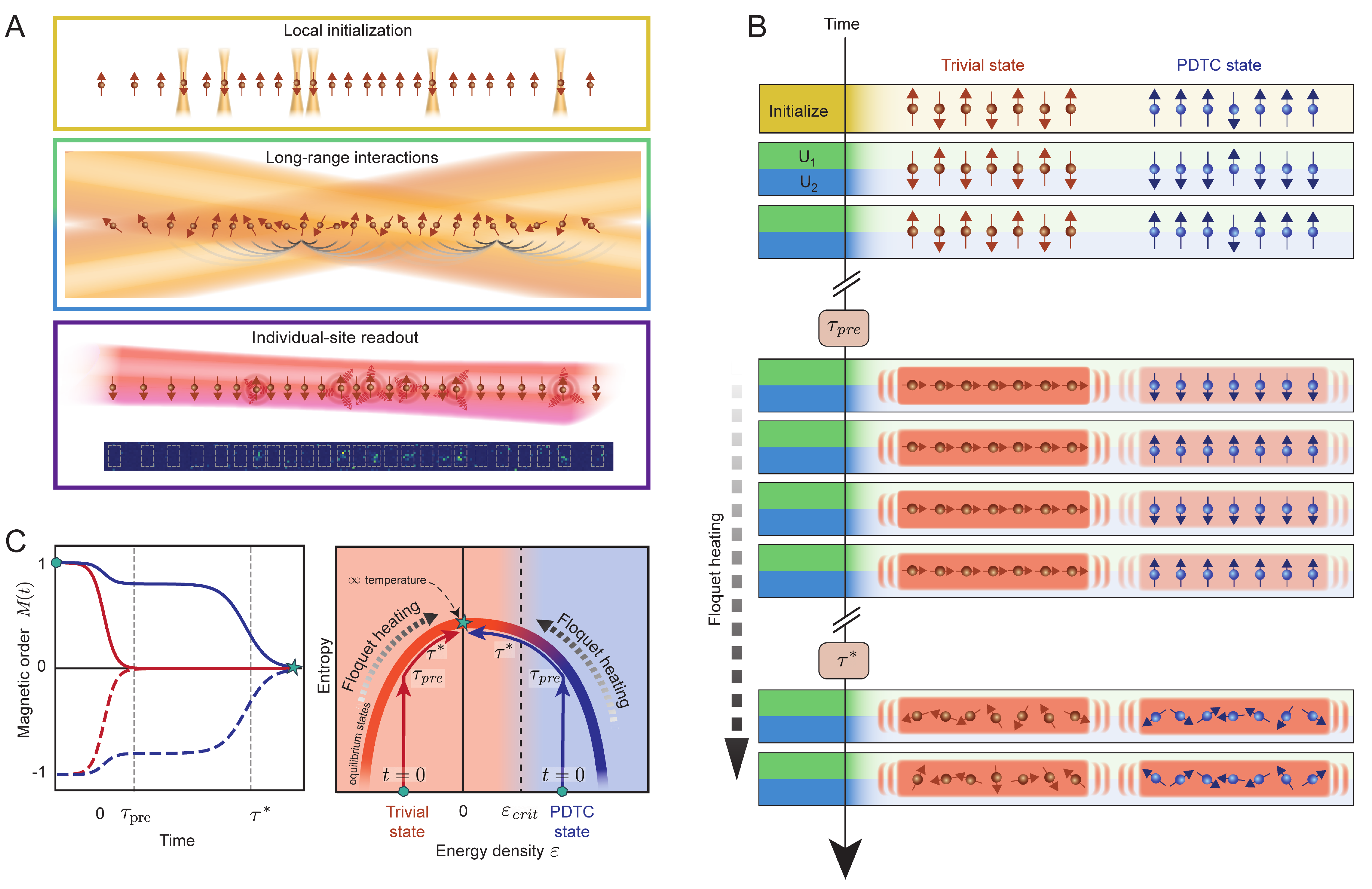}
\caption{\small
\textbf{Experimental setup and protocol}. 
\textbf{A.} Schematic of a linear array of 25 trapped atomic ion spins, fixed in space, as the physical platform for the experiments \cite{MonroeRMP2019}.
Top: single-site addressing enables the preparation of arbitrary product states of the spins.
Middle: two global Raman beams off-resonantly couple to motional modes and generate long-range Ising interactions.
Bottom: state-dependent fluorescence provides single-site readout, enabling the measurement of both magnetization and energy density.
\textbf{B.} For intermediate times between$\tpre$ and $\tstar$, the system approaches an equilibrium state of the prethermal Hamiltonian $\Heff$. 
After $\tpre$, the magnetization in the trivial Floquet phase remains constant. Meanwhile, in  the PDTC phase, the magnetization oscillates each period leading to a robust sub-harmonic response. For both phases, at times $t \gg \tstar$, Floquet heating eventually brings the many-body system to  a featureless infinite temperature ensemble. 
\textbf{C.} 
Left: Schematic of the stroboscopic magnetization dynamics in the trivial [red] and PDTC [blue] phase (full/dashed curves represent even/odd driving periods).
In the trivial phase, any transient time-crystalline-order decays by the prethermal equilibration time $\tpre$, while in the PDTC phase, the order remains robust until $\tstar$, the frequency-controlled heating timescale.
Right: Starting from a product state with zero entropy, the dynamics under $\Heff$ bring the system to an equilibrium state at time $\tpre$.
The PDTC behavior is robust if the initial state thermalizes to a prethermal equilibrium state, which spontaneously breaks an emergent symmetry of $\Heff$.
In our system, this occurs  if the  energy density of the initial state is above a critical value $\varepsilon_{\mathrm{crit}}$ (i.e.~to the right of the dashed line), wherein the system equilibrates to a ferromagnetic state.
Regardless of the initial state, for $t> \tstar$,  Floquet heating eventually brings the system to the maximum entropy state at zero energy density.
}
\label{fig:Fig1}
\end{figure*}



Recently, an alternate, disorder-free framework for addressing both challenges has emerged: Floquet \emph{prethermalization}~\cite{Kuwahara_Mori_Saito_2016, Abanin_DeRoeck_Ho_Huveneers_2017, Machado_Kahanamoku-Meyer_Else_Nayak_Yao_2019, Rubio-Abadal_Ippoliti_Hollerith_Wei_Rui_Sondhi_Khemani_Gross_Bloch_2020,Peng_2021}.
For sufficiently high Floquet drive frequencies, energy absorption by the many-body system requires multiple correlated local rearrangements, strongly suppressing the heating rate. 
The Floquet heating time $\tstar$ scales exponentially with the drive frequency and can thus be prolonged beyond experimentally practical timescales; of course, directly observing such an exponential is challenging in any experiment because of finite decoherence timescales.
For time $t < \tstar$, the system dynamics is then captured by an effective prethermal Hamiltonian $\Heff$ \cite{Kuwahara_Mori_Saito_2016, Abanin_DeRoeck_Ho_Huveneers_2017}. 
This prethermal Hamiltonian defines an effective energy for the Floquet system and also determines the nature of the prethermal state, which is reached at the much shorter local equilibration time $\tpre$. 
Thus, by focusing on times between $\tpre$ and $\tstar$, the dynamics are guaranteed to reflect the actual thermodynamic properties of the Floquet phase.

This intermediate prethermal regime is not necessarily trivial.
New symmetries, protected by the discrete time translation symmetry of the drive, can emerge and lead to intrinsically non-equilibrium phases of matter~\cite{Else_Bauer_Nayak_2017, Machado_Else_Kahanamoku-Meyer_Nayak_Yao_2020}.
One example of such a phase is the prethermal discrete time crystal (PDTC), in which the many-body system spontaneously breaks the discrete time translation symmetry of the drive and develops a robust sub-harmonic response.

A disorder-free PDTC exhibits a number of key differences  compared to the MBL discrete time crystal, despite the similarity of their sub-harmonic response \cite{Else2020,Khemani_Moessner_Sondhi_2019}. 
Stabilized by MBL, time-crystalline order is independent of the initial state and persists to arbitrarily late times, but is believed to only occur in low dimensions with sufficiently short-range interactions \cite{Yao_Laumann_Gopalakrishnan_Knap_Muller_Demler_Lukin_2014, DeRoeck_Huveneers_2017}.
By contrast, the PDTC lifetime is limited by  $\tstar$ and depends on the energy density of the initial state; this energy density determines the prethermal state to which the system equilibrates for times $t>\tpre$. 
Crucially, if the prethermal state spontaneously breaks an emergent symmetry of $\Heff$, then the many-body system will also exhibit robust time-crystalline order, corresponding to an oscillation between the different symmetry sectors \cite{Else_Bauer_Nayak_2017, Machado_Else_Kahanamoku-Meyer_Nayak_Yao_2020}.
On the other hand, if the prethermal state is symmetry-unbroken, any signatures of time-crystalline order will decay by $\tpre$ and the system is in a trivial Floquet phase. 

The energy density dependence of the PDTC phase can also be understood as the necessity for $\Heff$ to host a symmetry-breaking phase.
This again contrasts with the MBL discrete time crystal, because symmetry breaking is more easily realized in \emph{higher dimensions}.
Indeed, in one dimension, Landau-Peierls arguments rule out the existence of a PDTC with short-range interactions~\cite{Peierls_1936, Landau1937}, and long-range interactions are a necessary ingredient to stabilize a prethermal time crystal \cite{Machado_Else_Kahanamoku-Meyer_Nayak_Yao_2020}.

We exploit the controlled long-range spin-spin interactions of an ion trap quantum simulator to observe signatures of a one-dimensional prethermal discrete time crystal.
Our main results are three-fold. 
First, we prepare a variety of locally inhomogeneous initial states via individual addressing of ions within the one-dimensional chain (see Fig. \ref{fig:Fig1}A).
By characterizing the quench dynamics starting from these  states, we directly observe the approach to the prethermal state, enabling the experimental extraction of the prethermal equilibration time,  $\tpre$.
Second, we measure the time dynamics of the energy density as a function of the driving frequency. 
By preparing states at both positive and effectively negative temperatures (Fig. \ref{fig:Fig1}c), we observe either the gain or loss of energy as the system heats to infinite temperature (corresponding to zero energy density).
Importantly, we find that the heating timescale, $\tstar$, increases with the driving frequency (Fig.~2).
Finally, to probe the nature of prethermal time-crystalline order, we study the Floquet dynamics of initial states, spanning across the entire energy spectrum, that equilibrate to either a symmetry-breaking or a symmetry-unbroken ensemble. 
The former exhibits robust period-doubling behavior up until the frequency-controlled heating timescale, $\tstar$ (Fig.~3B).
In comparison, for the latter, all signatures of period doubling disappear by the frequency-independent timescale $\tpre$ (Fig.~3A).
By investigating the lifetime of the time-crystalline order as a function of the energy density of the initial state, we identify the phase boundary for the PDTC. This boundary is consistent with independent quantum Monte Carlo calculations for the location of the phase transition in $\Heff$. 

Our system consists of a one-dimensional chain of 25 $^{171}\textnormal{Yb}^+$ ions. Each ion encodes an effective spin-1/2 degree of freedom in its hyperfine levels $\ket{F=0,m_F=0}$ and $\ket{F=1,m_F=0}$ (Fig. \ref{fig:Fig1}a).
Long-range Ising interactions are generated via a pair of Raman laser beams~\cite{Molmer1999,Supp}.
Arbitrary effective magnetic fields can be applied either  locally or globally and single-site readout
can be performed simultaneously across the full chain \cite{MonroeRMP2019}, enabling the direct measurement of the Floquet dynamics of both the magnetization and the energy density. 

The Floquet drive alternates between two types of Hamiltonian dynamics (Fig. \ref{fig:Fig1}b): (i) a global $\pi$-pulse  around the $\hat{y}$ axis and (ii) evolution for time $T$ under a disorder-free, long-range, mixed-field Ising model. This is described by the two evolution operators,
\begin{eqnarray}\label{eq:FloquetDrive}
U_1 &=& \exp\left[i\frac{\pi}{2}\sum_i^N \sigma_i^y \right]\nonumber\\
U_2 &=& \exp \left[
iT\left( \sum_{i<j}^NJ_{ij}\sigma_i^x\sigma_j^x + \right.\right.\\
&& +\left.\left. B_y\sum_{i=1}^N\sigma_i^y + B_z\sum_{i=1}^N\sigma_i^z \right)\right],
\end{eqnarray}
where $\sigma_i^{v}$ is the $v$th component of the spin-1/2 Pauli operator for the $i$th ion, and we adopt the convention $\hbar=1$.
Here, $J_{ij} > 0$ is the long-range coupling with average nearest-neighbor interaction strength $J_0 = 2\pi\cdot0.33$ kHz, while $B_y=2\pi\cdot0.5$ kHz and $B_z=2\pi\cdot0.2$ kHz are global effective magnetic fields. 
The Floquet unitary $U_F = U_2U_1$ implements the dynamics over a period of the drive and has frequency $\omega=2\pi/T$.


Within the prethermal window in time $\tpre<t<\tstar$, the stroboscopic dynamics of the system (every other period) are well-approximated by an effective prethermal Hamiltonian, which to lowest order in $1/\omega$ is given by \cite{Machado_Else_Kahanamoku-Meyer_Nayak_Yao_2020}:
\begin{equation} \label{eq:H0}
\Heff = \sum_{i<j}^N J_{ij}\sigma_i^x\sigma_j^x + B_y\sum_{i=1}^N\sigma_i^y.
\end{equation}
We begin by characterizing the dynamics of the system as it approaches the prethermal state of $\Heff$.
In particular, we prepare an initial state with all spins pointing along $\hat{x}$ (in an eigenstate of $\sigma^x$), except for two central spins, which are prepared along $\hat{z}$ (Fig. 2D).
Quench dynamics from this initial state show that the  $\hat{x}$-magnetization of the two central spins exhibits two-step dynamics. The magnetization first equilibrates to the value of the neighboring spins, before decaying back to zero at late times. 
As depicted in Fig.~2D, the convergence of the initially inhomogeneous magnetization to a uniform finite value demonstrates that the system first reaches an intermediate-time equilibrium (i.e.~prethermal) state before ultimately heating to infinite temperature.
We find that this prethermal timescale is approximately given by $J_0\tpre \approx 3$. 

\begin{figure*}
\centering
\includegraphics[width=4.2in]{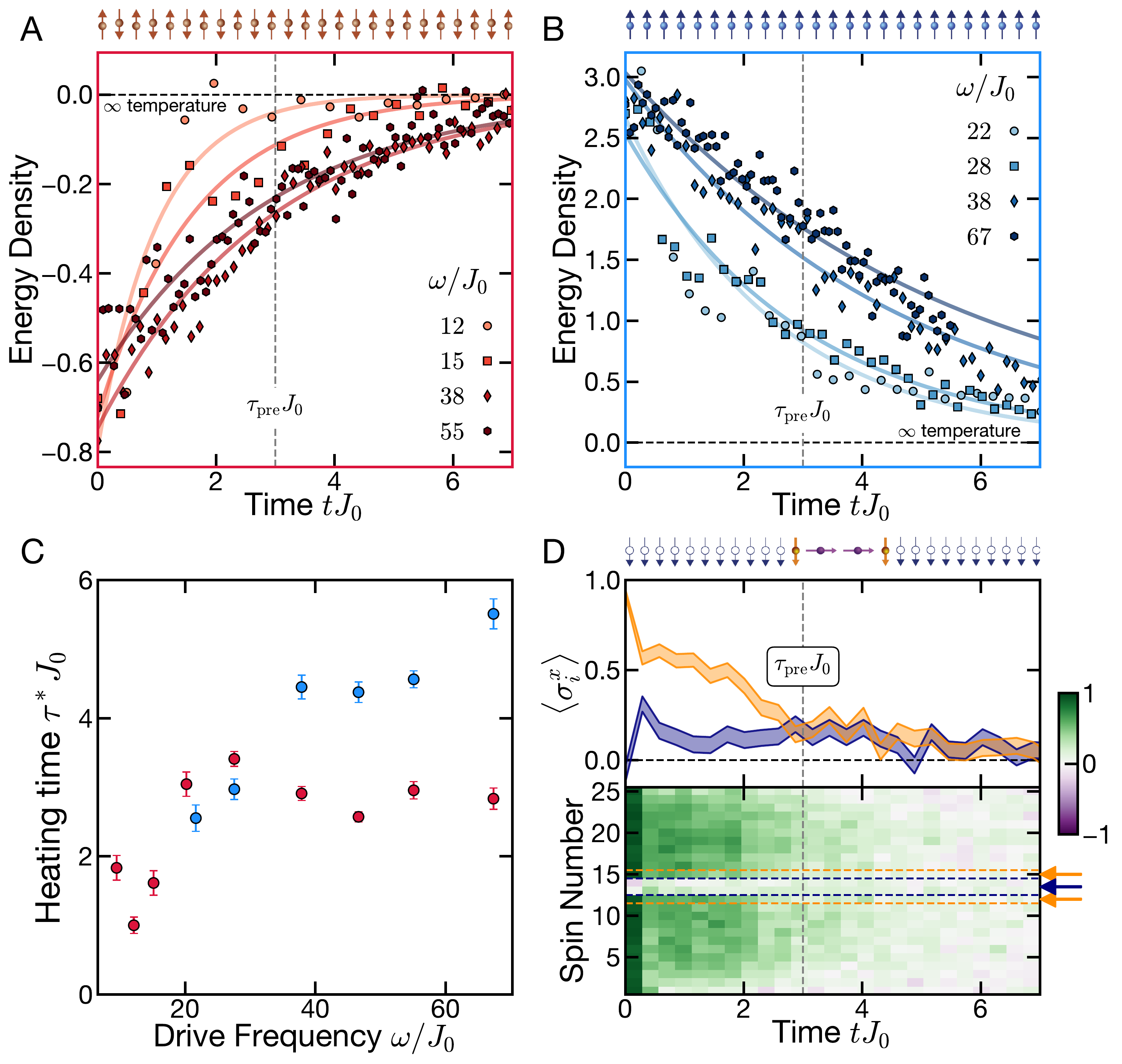}
\caption{\small
\textbf{Characterizing the prethermal regime.}
\textbf{A, B. }
The dynamics of the energy density for a low-energy N\'eel state (left) and a high-energy polarized state (right).
Both states exhibit Floquet heating toward infinite temperature, albeit from opposite sides of the many-body spectrum. 
In addition, in both cases, faster drive frequencies $\omega$ suppress the heating rate. 
Statistical error bars are of similar size as the point markers.
\textbf{C. } 
The heating time $\tstar$ for the N\'{e}el (red) and polarized (blue) states increases with frequency. 
$\tstar$ is extracted by fitting the dynamics of the energy density to $\sim e^{-t/\tstar}$; exponential fits are shown as solid curves in (A) and (B).
At high frequencies this behavior saturates owing to the presence of external noise. Error bars for the heating time correspond to fit errors.
\textbf{D. }
The prethermal equilibration time, $\tpre$, can be characterized by observing the local $\hat{x}$-magnetization dynamics for even Floquet periods. 
Top: The middle two spins (purple), initially prepared along a different axis, rapidly align with their neighbors (orange) at time $\tpre J_0\approx3$.
The homogenization of the magnetization signals local equilibration to a equilibrium state of $\Heff$.
Bottom: The magnetization of each ion as a function of time.
}
\label{fig:Fig2}
\end{figure*}

In addition to $\tpre$, the prethermal regime is also characterized by the timescale associated with the frequency-dependent Floquet heating, $\tstar$.
To experimentally investigate $\tstar$, we measure the dynamics of the prethermal energy density, $\langle \Heff \rangle/(N J_0)$, for two different initial states on opposite ends of the many-body spectrum of $\Heff$: a low-energy N\'{e}el state (Fig.~2A) and a high-energy polarized state (Fig.~2B) along the Ising interaction direction.
In both cases, we observe the expected trend: increasing the driving frequency suppresses the  heating rate (Fig.~2C).
For sufficiently large frequencies, we observe a plateau in $\tstar$, suggesting the presence of  external noise. In particular, the origin of this plateau is consistent with experimental fluctuations of the light intensity, leading to noisy AC Stark shifts on the qubit transition. A detailed analysis and numerical simulations of this noise are presented in the supplementary information~\cite{Supp}.

Our demonstration of the frequency dependence  of $\tstar$ (Fig.~2) directly translates into our ability to control the lifetime of the prethermal time crystal. 
As aforementioned, the key ingredient underlying  time-crystalline order is  the presence of an emergent symmetry, $G$, in $\Heff$.
Crucially, for a single period of evolution, the  exact Floquet dynamics are approximately generated by evolving under $\Heff$ for time $T$, followed by  acting with $G$ \cite{Else_Bauer_Nayak_2017, Machado_Else_Kahanamoku-Meyer_Nayak_Yao_2020}.
In our experiment, this symmetry corresponds to a global spin flip, $G\approx U_1 \propto \prod_{i=1}^N \sigma^y_i$.
This implies that the time-crystalline order is naturally captured by the system’s magnetization dynamics, which flips the order parameter $\langle \sigma^x_i \rangle$.
As a result, there are two possibilities for the prethermal dynamics (Fig.~1B).
If the prethermal state respects the symmetry, the magnetization is zero and remains unchanged across a period.
Conversely, if the prethermal state spontaneously breaks the symmetry, the magnetization is non-zero and  alternates every period. The resulting $2T$-periodic, sub-harmonic dynamics is the hallmark of a time crystal.

Long-ranged ferromagnetic Ising interactions stabilize a low-temperature one-dimensional ferromagnetic phase but long-ranged Ising interactions that are antiferromagnetic between all pairs of spins do not stabilize a low-temperature one-dimensional antiferromagnetic phase. Thus, the antiferromagnetic interactions in our system do not support a symmetry-breaking phase at low energy densities but rather at the top of the spectrum at high densities (Fig.~1C), which is the low-energy-density regime of the ferromagnetic Hamiltonian $-\Heff$.
As a result, whether we prepare a low or high energy density initial state dictates whether the prethermal dynamics are trivial or in the PDTC phase.

\begin{figure*}
\centering
\includegraphics[width=\textwidth]{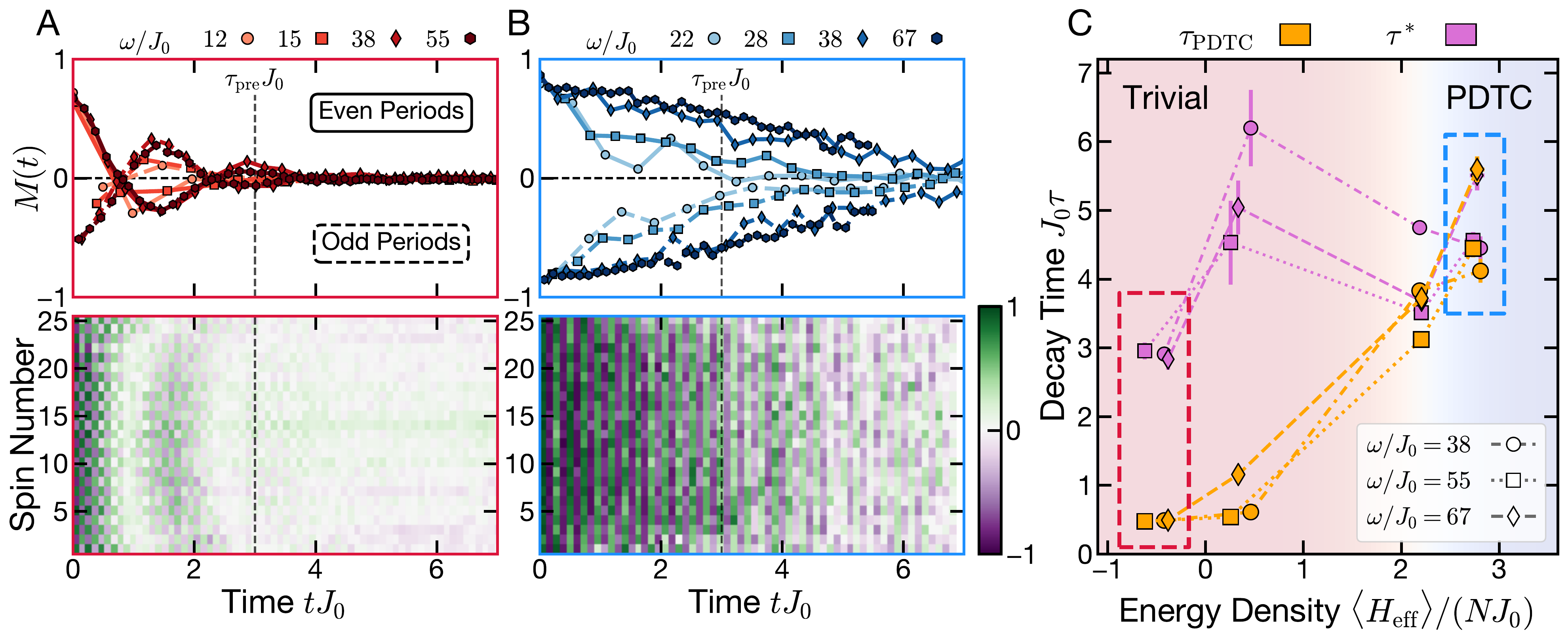}
\caption{\small\textbf{Characterizing the PDTC phase. }\textbf{A, B.} 
Upper plots: The magnetization dynamics, $M(t)$, for the N\'{e}el state (left) and the polarized state (right). 
For the N\'{e}el state, $M(t)$ quickly decays to zero at time $\tpre$ (dashed vertical line), independent of the drive frequency. 
For the polarized state, the sub-harmonic response (2$T$-periodicity) persists well-beyond $\tpre$ and its lifetime is extended upon  increasing  the drive frequency.
We characterize the lifetime of the prethermal time-crystalline order by fitting the magnetization dynamics to an exponential and extracting a decay time, $\tTC$ \cite{Supp}.
Statistical error bars are of similar size as the point markers.
Lower plots: The $\hat{x}$-magnetization dynamics for each ion in the chain at $\omega/J_0=38$. 
\textbf{C.} Heating ($\tstar$) and magnetization decay ($\tTC$) times for four different initial states at varying energy densities \cite{Supp}. 
For low energy densities, $\tTC$ (orange) are short, independent of frequency, and significantly shorter than $\tstar$ (magenta), highlighting the trivial Floquet phase. 
For high energies, $\tTC$ is similar to  $\tstar$, highlighting the long-lived, frequency-controlled nature of the PDTC behavior.
The location of the observed crossover in energy density is in agreement with an independent quantum Monte Carlo calculation (red and blue shaded regions) \cite{Supp}.
Error bars for the decay time correspond to fit errors, while  error bars for the energy density  correspond to statistical errors.}
\label{fig:Fig3}
\end{figure*}

We investigate these two regimes by measuring the auto-correlation of the magnetization: 
\begin{eqnarray}
M(t) = \frac{1}{N}\sum_{i=1}^{N}\ensav{\sigma_i^x(t)}\ensav{\sigma_i^x(0)}.
\end{eqnarray}
Starting with a low-energy-density N\'{e}el state (Fig.~3A), we observe that $M(t)$ quickly decays to zero at $\tpre$, in agreement with the expectation that the system equilibrates to the symmetry-unbroken, paramagnetic phase.
This behavior is frequency-independent, in direct contrast to the Floquet dynamics of the energy density (Fig.~2A).
This contrast highlights an essential point: although $\tstar$ can always be extended by increasing the driving frequency, if the system lives in the trivial Floquet phase, no order will survive beyond $\tpre$.

The Floquet dynamics starting from the polarized state are markedly distinct (Fig.~3B).
First, $M(t)$ exhibits period doubling, with $M>0$ for even periods and $M<0$ for odd periods.
Second, the decay of this period-doubling behavior is directly controlled by the frequency of the drive. 
Third, the lifetime of the time-crystalline order  mirrors the dynamics of the energy density shown in Fig.~2B, demonstrating that Floquet heating ultimately melts the PDTC at late times.

By considering two additional initial states, we explore the stability of the PDTC phase as a function of  energy density.
Fig.~3C depicts both the heating time as well as the lifetime of the time-crystalline order. 
Near the  bottom of the spectrum, where no symmetry-breaking phase exists, the decay of the magnetization is both frequency-independent and significantly faster than the heating timescale.
By contrast, near the top of the spectrum, where the symmetry-breaking ferromagnetic phase lies, the two timescales are consistent with one another, demonstrating that the PDTC lifetime is limited by  Floquet heating. 
Our results are consistent with a phase boundary occurring around energy density $\langle \Heff\rangle/ (NJ_0) \approx 2$, in agreement with independent numerical calculations via quantum Monte Carlo  \cite{Supp}.

In this work we report the experimental observation of robust prethermal time-crystalline behavior that persists beyond any early-time transient dynamics.
By varying the energy density of the initial state, we study the crossover between the trivial Floquet phase and the PDTC phase.
Our results highlight the potential of periodic driving, in general, and prethermalization, in particular, as a framework for realizing and studying out-of-equilibrium phenomena. 
Even in the presence of noise, we find that the prethermal dynamics remain stable,  suggesting the possibility that an external bath at sufficiently low  temperature can stabilize the prethermal dynamics for infinitely long times~\cite{Else_Bauer_Nayak_2017}.
This stands in contrast to localization-based approaches for stabilizing Floquet phases, in which the presence of an external bath tends to destabilize the dynamics.
Our work opens the door to a number of intriguing future directions: (i) exploring generalizations of Floquet prethermalization to a quasi-periodic drive\cite{Else_Ho_Dumitrescu_2020}, (ii) stabilizing Floquet topological phases~\cite{Potirniche_2017,Else_Fendley_2017}, and (iii) leveraging non-equilibrium many-body dynamics for enhanced metrology \cite{Choi_Yao_Lukin_2017}.

\section*{Acknowledgments}
We acknowledge fruitful discussions with C. Laumann, W. L. Tan, A. Vishwanath, D. Weld, and J. Zhang. 
\textbf{Funding:} This work is supported by the DARPA Driven and Non-equilibrium Quantum Systems (DRINQS) Program D18AC00033, NSF Practical Fully-Connected Quantum Computer Program PHY-1818914, the DOE Basic Energy Sciences: Materials and Chemical Sciences for Quantum Information Science program DE-SC0019449, the DOE High Energy Physics: Quantum Information Science Enabled Discovery Programs DE-0001893,  the AFOSR MURI on Dissipation Engineering in Open Quantum Systems FA9550-19-1-0399, the David and Lucile Packard foundation, the W.~M.~Keck foundation, and the EPiQS Initiative  of  the  Gordon  and  Betty  Moore  Foundation GBMF4303.
\textbf{Author contributions:} A.K., W.M., P.B., K.S.C., L.F., P.W.H., G.P., and C.M. designed and performed experimental research, F.M., D.V.E., C.N., and N.Y.Y. analyzed the data theoretically, and all authors wrote the paper.
\textbf{Competing interests:} C.M. is the co-founder and Chief Scientist at IonQ, Inc.
\textbf{Data availability:} All data needed to evaluate the conclusions in the paper are present in the paper. Additional data related to this paper may be requested from the corresponding author.

\setcounter{figure}{0}
\makeatletter 
\renewcommand{\thefigure}{S\@arabic\c@figure}
\makeatother


\clearpage

\section*{Supplementary materials}

\subsection*{The trapped-ion quantum simulator}

 The quantum simulator used in this work is based on a chain of $^{171}$Yb$^+$ ions trapped in a 3-layer Paul trap at room temperature \cite{Kihwan2009}.
The ions are confined in all three directions by a combination of static and oscillating electric fields. 
The interplay of the repulsive Coulomb force and the trapping potential arrange the ions in a linear configuration.
The transverse center-of-mass (COM) motional mode frequency along the $\hat{x}$ axis is $f_\textnormal{COM} = 4.67 $ MHz and the axial COM frequency is $f_z = 0.34 $ MHz. The axial trapping strength is set such that 25 ions settle in a linear configuration with inter-ion spacings varying from $\sim 2~\mu$m (chain center) to $\sim 3.5~\mu$m (chain edges) \cite{James1998}.

\subsection*{Spin and motional state preparation}

Between experiments, the ions are Doppler cooled by a 369.5 nm laser red-detuned from the ${}^2S_{1/2}$ to ${}^2P_{1/2}$ transition by 10 MHz, one-half of the transition linewidth. This laser projects onto all three principle axes of the trap, ensuring that the ions are cooled along all directions. To begin an experiment, the ions are initialized in the low-energy hyperfine qubit state \mbox{$\down \equiv {}^2S_{1/2} \ket{F = 0, m_F = 0}$} by an incoherent optical pumping process \cite{Olmschenk2007}. Optical pumping requires approximately $20 ~\mu$s and initializes all ions to $\down$ with at least 99~\% fidelity. At this point the individual spin states of the ions are well-known, while the shared motional state is a thermal distribution with $\bar{n} \leq 3$ average motional quanta in the transverse $\hat{x}$ axis modes. Resolved sideband cooling on multiple motional modes brings the ions near their motional ground state ($\bar{n}\leq 0.1$ average motional quanta).

With the ions cooled and their spin states initialized, we prepare the spins in product states along the $\hat{x}$ axis of the Bloch sphere with a combination of global rotations and individual $\sigma^z$ rotations. Global rotations are driven with a pair of Raman laser beams, intersecting at a $90^{\circ}$ angle. These lasers produce a beatnote that drives oscillations between the qubit states with Rabi frequency $\Omega$ when tuned on resonance with the ions' S-manifold hyperfine splitting. The phase of this beatnote determines the Bloch sphere axis about which the spins are rotated. 

\subsection*{The BB1 pulse sequence}
Each Raman beam has a Gaussian intensity profile with waists of $10~\mu$m by $130~\mu$m at the ion plane. A 25-ion chain has a length of about $60~\mu$m. Each ion samples a slightly different intensity from the Raman lasers, resulting in different rates of rotation across the chain. To minimize rotation errors caused by this inhomogeneity, we employ BB1 dynamical decoupling sequences\cite{BB1} to ensure that all spins along the chain are rotated by the same amount.

A traditional $\hat{y}$ rotation unitary has the form $\hat{U}^y_\theta = e^{-i \theta\sigma^y_i/2}$, where $\theta$ is the desired angle of rotation about the $\hat{y}$ axis. The angle $\theta \equiv \Omega_i t$, where $\Omega_i$ is the Rabi frequency experienced by spin $i$, is sensitive  to the spacially-inhomogeneous intensity profiles of the Raman lasers. We instead apply the following BB1 unitary, consisting of 4 sub-rotations:
\begin{equation}
    \hat{U}^y_{\theta,\textnormal{BB1}} = e^{-i \frac{\pi}{2}\sigma^\phi_i} e^{-i \pi\sigma^{3\phi}_i} e^{-i \frac{\pi}{2}\sigma^\phi_i} e^{-i \frac{\theta}{2}\sigma^y_i}.
\end{equation}
The phase $\phi$ depends on the desired rotation angle $\theta$:
\begin{equation}
    \phi = \arccos\left( \frac{\theta}{4 \pi} \right).
\end{equation}
While a $\pi$-rotation using this sequence takes five times longer than a traditional rotation pulse, it reduces rotation errors significantly and prevents dephasing across the chain. This allows us to apply hundreds of $\pi$-pulses with negligible loss of contrast---a requirement for the time-crystal experimental sequences presented in this manuscript.

\subsection*{Arbitrary product state preparation}

An individual addressing beam focused to a waist of $500$ nm generates rotations on each spin with relatively low crosstalk. A high-bandwidth acousto-optical deflector (AOD) steers the beam, and the AOD's rf drive frequency maps the beam to a location along the ion chain. This beam applies a fourth-order AC Stark shift to the hyperfine qubit splitting\cite{Lee2016}, creating an effective $\sigma^z_i$ rotation on a single spin $i$. This rotation is mapped to a rotation about any axis using the appropriate global analysis $\pi/2$ rotations, allowing for preparation of product states with arbitrary spin flips such as the antiferromagnetic N\'{e}el state.

\subsection*{Qubit readout}

At the end of an evolution, we measure the magnetization of each spin using state-dependent fluorescence. A 369.5 nm laser resonant with the ${}^2S_{1/2}\ket{F=1} \leftrightarrow {}^2P_{1/2}\ket{F=0}$ transition (linewidth $\gamma/2\pi \approx 19.6$ MHz) causes each ion to scatter photons if the qubit is projected to the $\up$ state. Ions projected to the $\down$ qubit state scatter a negligible number of photons because the laser is detuned from resonance by the ${}^2S_{1/2}$ hyperfine splitting. By applying global $\pi/2$-rotations, we rotate the $x$ and $y$ bases into the $z$ basis. This allows us to measure all individual magnetizations and many-body correlators along any single axis. In the experiments reported in this work, we repeat the experimental sequence and the measurement for 50-600 times to reduce quantum projection noise.

For each measurement, a finite-conjugate $N\!A=0.4$ objective lens system (total magnification of $70\times$) collects scattered 369.5 nm photons and images them onto an Andor iXon Ultra 897 EMCCD camera. Before taking data, high-contrast calibration images of the ion chain, illuminated by Doppler cooling light, are used to identify a region of interest (ROI) on the camera sensor for each ion. During data collection, fluorescence is integrated for $400~\mu$s, after which a pre-calibrated binary threshold is applied to discriminate the qubit state of each ion with approximately 98~\% accuracy per ion. 

The dominant error sources for the qubit readout, ordered by decreasing significance, are: mixing of qubit states caused by off-resonant coupling during the $400~\mu$s camera exposure window, crosstalk between ion ROIs due to small inter-ion spacings near the center of the chain, electronic camera noise, and laser power fluctuations. No state preparation and measurement (SPAM) correction has been applied to data presented in this work.

\subsection*{Simulating the transverse field Ising Hamiltonian}
We generate the effective spin-spin interaction Hamiltonian by applying spin-dependent dipole forces with the pair of 355 nm Raman beams mentioned earlier. These beams produce a beatnote with wavevector $\Delta \vec{k}$ aligned along a principle axis of the trap. The frequencies of these beams are controlled with acousto-optical modulators (AOMs) to generate a pair of beatnote frequencies detuned by $-\mu$ (red beatnote) and $+\mu$ (blue beatnote) from the resonant qubit transition frequency. For $\mu - f_{COM} \gg \eta\Omega$ ($\eta$ is the Lamb-Dicke parameter\cite{NistBible}) and $\eta \ll 1$, the experiment operates in the far-detuned M\o{}lmer-S\o{}rensen (MS) regime\cite{Molmer1999,Kihwan2009}. Here, excited motional states are adiabatically eliminated and the laser-ion interaction takes the form of a spin-spin, effective long-range interacting Ising Hamiltonian
\begin{equation}
    H = \sum^N_{i<j} J_{ij} \sigma^x_i \sigma^x_j.
    \label{eq:Ising}
\end{equation}
The $N \times N$ matrix $J_{ij}$ describes couplings between spins $i$ and $j$ (Fig.~S1):
\begin{equation}
    J_{ij} = \Omega^2 f_R \sum^N_{m=1} \frac{b_{im}  b_{jm} }{\mu^2-f^2_m}
\end{equation}
where $f_R = {\Delta k}^2/(2 M)$ is the recoil frequency, $f_m$ is the frequency of the $m$-th motional mode, $b_{im}$ is the eigenvector matrix element of the $i$-th ion's participation in the $m$-th motional mode ($\sum_i|b_{im}|^2=\sum_m|b_{im}|^2 = 1$), and $M$ is the mass of a single ion.
Using $\eta_m=\sqrt{f_R/f_m}$, we get
\begin{equation}
J_{ij} \approx \frac{\Omega^2}{2}\sum_{m=1}^N\frac{b_{im}  b_{jm}\eta_m^2}{\delta}
 \label{JijInteraction}
\end{equation}
\begin{figure*}[t]
    \centering
    \includegraphics[width=\textwidth]{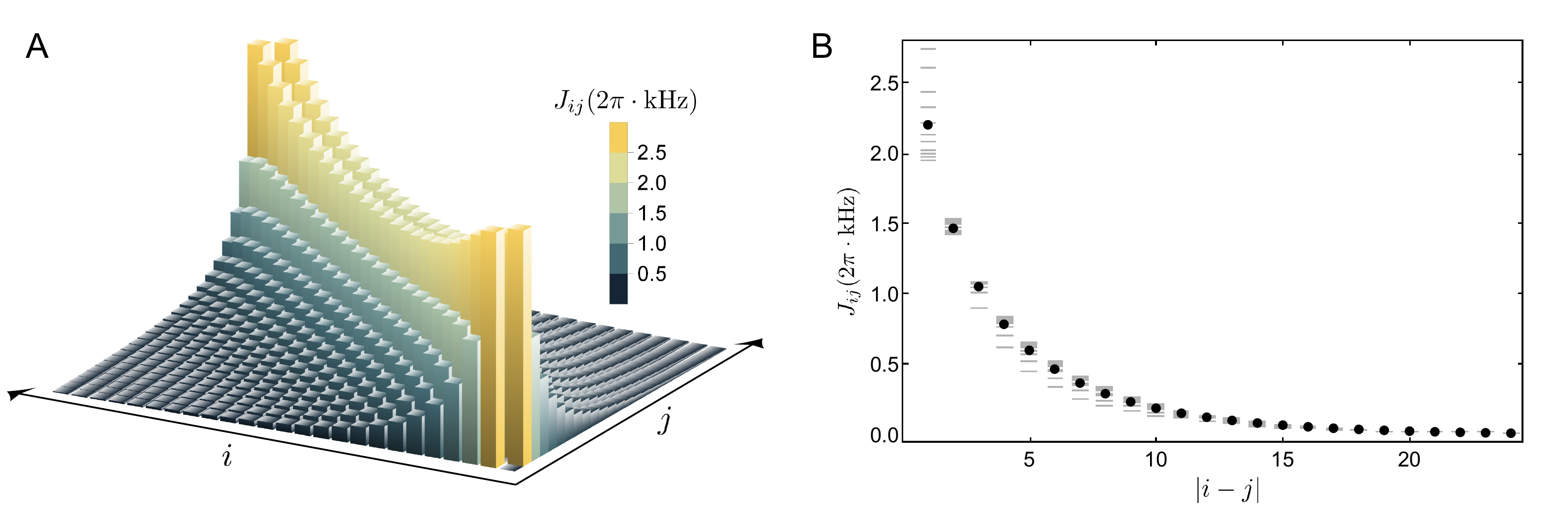}
    \caption{\small\textbf{The interaction matrix $J_{ij}$.} \textbf{A.} The position of each column represents a pair of spins $\{i,j\}$, while its color-coded height the strength of their coupling.
    \textbf{B.} To illustrate how the interaction strength scales with distance between $i$ and $j$, we average the elements of the diagonals of $J$ (circular markers), e.g.~the $|i-j|=2$ corresponds to $\ensav{\{J_{13},J_{24},J_{35},J_{46},\ldots\}}$. The line markers illustrate all the individual couplings for that value of $|i-j|$.
    }
    \label{fig:JijPlots}
\end{figure*}
In the reported experiments, the average nearest-neighbor interaction $J_{0}/(2\pi) = 0.33\pm0.02$ kHz. 

We add transverse fields to Eq.~\ref{eq:Ising} in two ways. To create an effective transverse field $B_z$ along the $\hat{z}$ direction, we apply a global offset of $2B_z$ to the two Raman beatnote frequencies, imposing a rotating frame shift between the qubit and the beatnotes to generate an effective field with strength $B_z$. A third Raman beatnote, resonant with the qubit transition and applied simultaneously with the M\o{}lmer-S\o{}rensen red and blue beatnotes, creates additional transverse fields along the $\hat{x}$ or $\hat{y}$ directions depending on the beatnote phase. Altogether, the long-range, transverse-field Ising Hamiltonian takes the form

\begin{equation}
    H = \sum^N_{i<j} J_{ij} \sigma^x_i \sigma^x_j + B_y\sum^N_i \sigma^y_i + B_z\sum^N_i \sigma^z_i.
\end{equation}

\subsection*{Tukey window pulse shaping}
All Raman laser operations and individual-addressing operations are implemented via amplitude modulation of the rf that drives various AOMs and AODs. An arbitrary-wavefrom generator (AWG) outputs this amplitude-modulated rf signal to switch lasers and beatnotes on and off according to the experimental sequence. If the control rf is modulated with square pulses, the sharp edges (limited by the rise/fall time of the AOM/D) cause significant spectral broadening of the signal in the Fourier domain. This effect is more pronounced for shorter pulses, such as the Raman pulses used to generate the unitary $U_2$. This spectral broadening can be on the order of MHz, and causes undesirable driving of qubit motional and spin transitions.

We suppress spectral broadening by applying Tukey window pulse shaping, where the first and last 10 $\mu$s of the pulse are multiplied respectively by a rising and falling sinusoidal envelope. We account for the resulting reduction in the magnitude of each term of the Hamiltonian by scaling it down by an appropriate factor, which varies with the total duration of that pulse. Longer pulses need to be scaled down less, since the ramp time is fixed.

\subsection*{Error sources}

The dynamics observed in this work are the combination of ideal Hamiltonian evolution as in Eq.~\ref{eq:FloquetDrive} and other terms of smaller magnitude that we refer to as ``error sources''. The combined effect of the latter, when measuring the chain magnetization, manifests as decoherence.

The most significant error source is fluctuating AC Stark shifts of the hyperfine qubit frequency. This fluctuation is mostly caused by power instability of the 355 nm laser light at the ions' location. Even though there is a power PI locking scheme in effect for the 355 nm light, the sampling point for the lock is at a more upstream location that the ions. As the beams propagate downstream from that point, active elements, acoustic noise, and air turbulence introduce extra power noise. At the ions location, the light's red and blue beatnotes ideally produce exactly opposite AC Stark shifts of the qubit levels and cancel each other, in practice they are not always perfectly balanced. In this case, common-mode power fluctuations will make the sum of their Stark shifts fluctuate. This manifests as an effective fluctuating magnetic field term $B^{(AC)}(t)\sum_i\sigma_i^z$, common for all spins $i$ and is present in every stage of the experimental sequence. A fortunate side effect of the $\pi$-rotations of the drive is that they echo out part of this noise. However, the spectral portion of $B^{(AC)}(t)$ that is faster than $\omega/2$ is not echoed out and differs between different repetitions, manifesting as decoherence in the final averaged signal. In numerics presented in the next section, we model this noise based on experimental evidence and reasonable simplifications, and present numerical simulations that include it.

Imperfect qubit state readout also impacts the final fidelity of the simulation. During the finite readout window of $400 \mu$s, there is a small probability that a $\down$ state will be off-resonantly pumped, and read out as a $\up$ state, and vice versa. For the experiments presented in this work, the average readout error was $2.3\%$ for each ion.

Another error source comes from a term combining the spin and the motional part of the qubit wavefunction that acts in parallel with the effective Ising interaction in Eq.~(\ref{eq:Ising}). This term represents entanglement between these two parts; when we measure the qubit spin, we effectively trace out the entangled motional state, resulting in a probabilistic mixed state.  The probability for such an erroneous spin flip to occur is proportional to
\begin{equation}\label{eq:SpinFlips}
\sum_{m=1}^N \left(\frac{\eta_{m}b_{im}\Omega}{\delta_m}\right)^2
\end{equation}
Therefore, by increasing this detuning, we minimize the undesired spin-motion entanglement, but we are also decreasing the strength of our spin-spin interaction term (see Supplement, Simulating the transverse field Ising Hamiltonian). We set the balance between these effects by keeping the sum in Eq.~(\ref{eq:SpinFlips}) less than $0.1$ for two spins, which for the 25 spins results in approximately $0.7\%$ flip probability per spin. This effect is somewhat amplified by the finite duration of the Hamiltonian quenches in the second term of the Floquet drive, whose spectral decomposition has nonzero components in the motional frequencies. We considerably mitigate this effect by applying the Tukey window shaping to the relevant pulses (see Supplement, Tukey window pulse shaping), which reduces these undesired spectral terms.


\section*{Supporting numerical evidence}


\subsection*{Dynamics in the absence and presence of noise}

In this section, we present a numerical investigation of the dynamics simulated by the experimental platform, highlighting the physics of prethermalization and the PDTC.
We first focus on the noiseless case, where we observe the expected exponential frequency dependence of $\tau^*$ on the drive frequency $\omega$, and then turn to studying the effect of noise on the observed dynamics.

\begin{figure*}
    \centering
    \includegraphics[width =\textwidth]{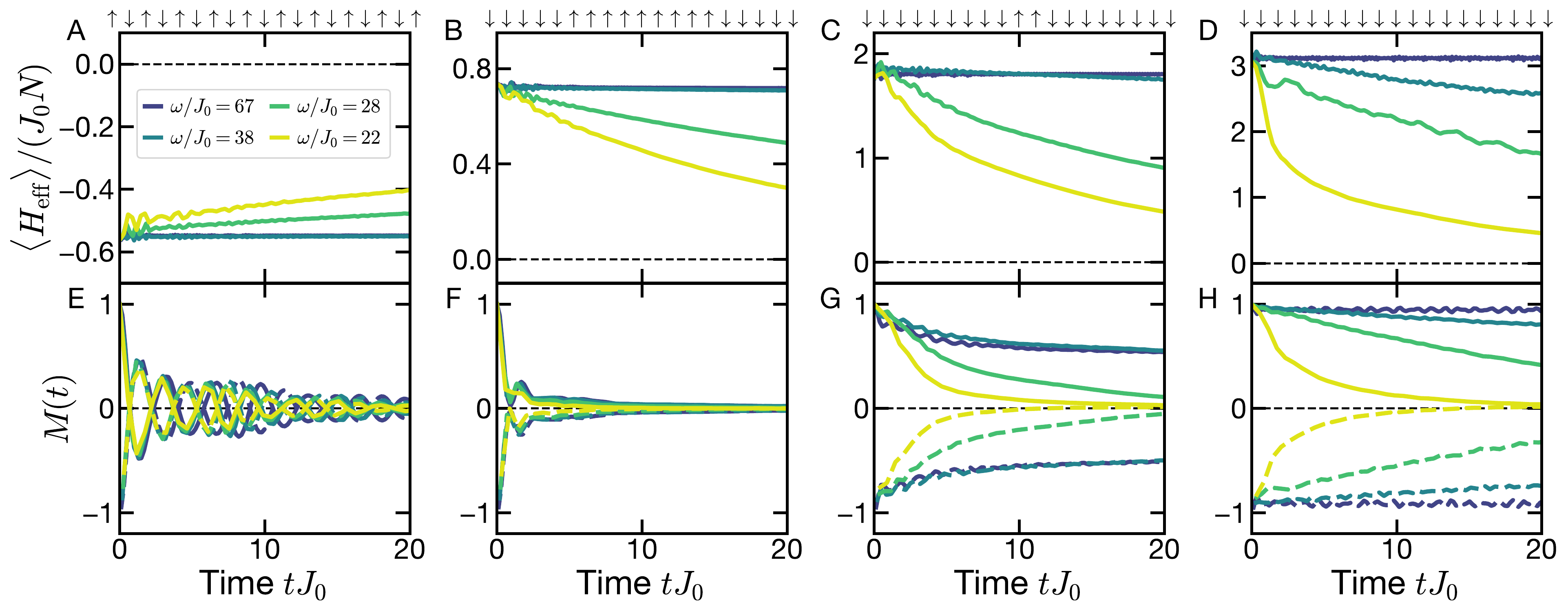}
    \caption{\small
    \textbf{\textbf{PDTC dynamics in the absence of noise}} 
    \textbf{A-D.} Dynamics of the energy density $\langle \Heff\rangle / NJ_0$ for different initial states with increasing energy density. 
    The decay of the energy density to the late time infinite temperature value is exponentially sensitive to the frequency of the drive.
    \textbf{E-H.} Dynamics of the magnetization $M(t)$, for different initial states. Full[dashed] line corresponds to even[odd] periods. 
    For low energy density, the magnetization is approximately frequency-independent, highlighting the trivial nature of the dynamics.
    For high energy density, the magnetization decay follows the decay of the energy density and exhibits a robust period doubling behavior---the two hallmarks of the PDTC.
    }
    \label{fig:NoNoise}
\end{figure*}

We consider an $N=19$ spin chain with interactions given by the experimentally determined interaction matrix for 25 spins, truncated to the middle $19$ spins.
We perform the exact time evolution of the full quantum system using Krylov subspace methods, simulating the entire experimental protocol (which includes the Tukey window pulse shaping).

\begin{figure*}
    \centering
    \includegraphics[width =0.5\textwidth]{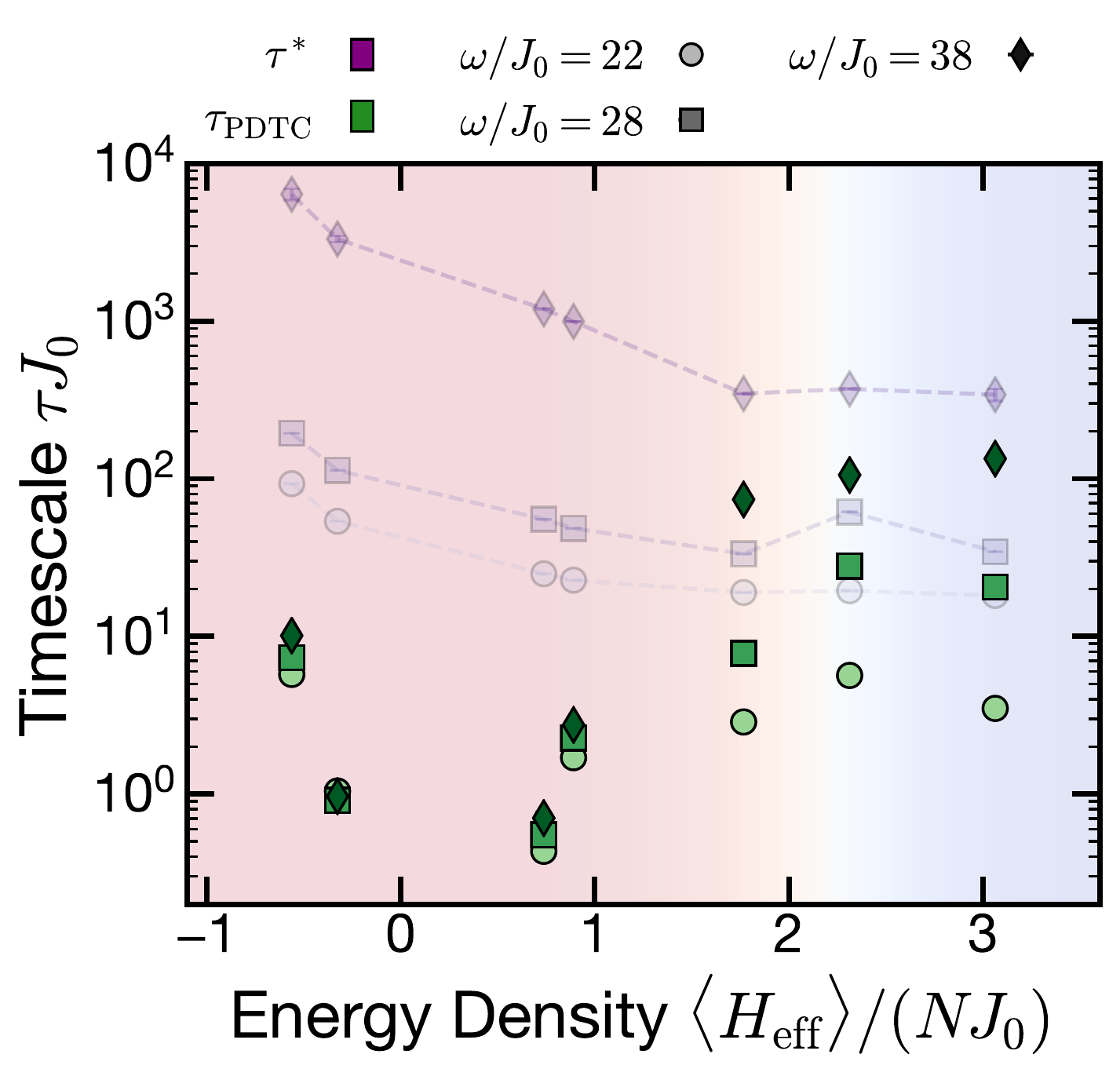}
    \caption{\small
    \textbf{PDTC  timescales in absence of noise}.
    Summary of the decay time scales of energy density ($\tau^*$) and magnetization ($\tau_{\mathrm{PDTC}}$).
    In the region where $\Heff$ exhibits a spontaneous symmetry broken phase (blue shading), $\tau_{\mathrm{PDTC}}$ follows the frequency dependence of $\tau^*$.
    By contrast, outside this region, the $\tau_{\mathrm{PDTC}}$ is much smaller than $\tau^*$ and is mostly frequency-independent.    
    }
    \label{fig:NoNoiseTimes}
\end{figure*}

For all states considered (Fig.~\ref{fig:NoNoise}), we observe the same effect of the frequency on the heating timescale $\tau^*$---the larger the frequency, the slower the energy approach to its infinite temperature value.
By contrast, the dynamics of the magnetization can be starkly different.
For the states at low enough energy density, where $H_{\mathrm{eff}}$ does not exhibit a spontaneous symmetry-breaking phase, the dynamics of $M(t)$ are mostly frequency-independent, and much faster than the heating.
For states near the top of the spectrum, where a spontaneous symmetry-breaking phase exists, the dynamics of the magnetization exhibit robust period doubling whose decay matches that of the energy density---this is PDTC behavior.
We summarize the timescales observed in Fig.~\ref{fig:NoNoiseTimes}, where the frequency dependence of $\tau^*$ occurs across all initial states, while the frequency dependence of $\tau_{\mathrm{TC}}$ only occurs at the top of the spectrum.

\begin{figure*}
    \centering
    \includegraphics[width =\textwidth]{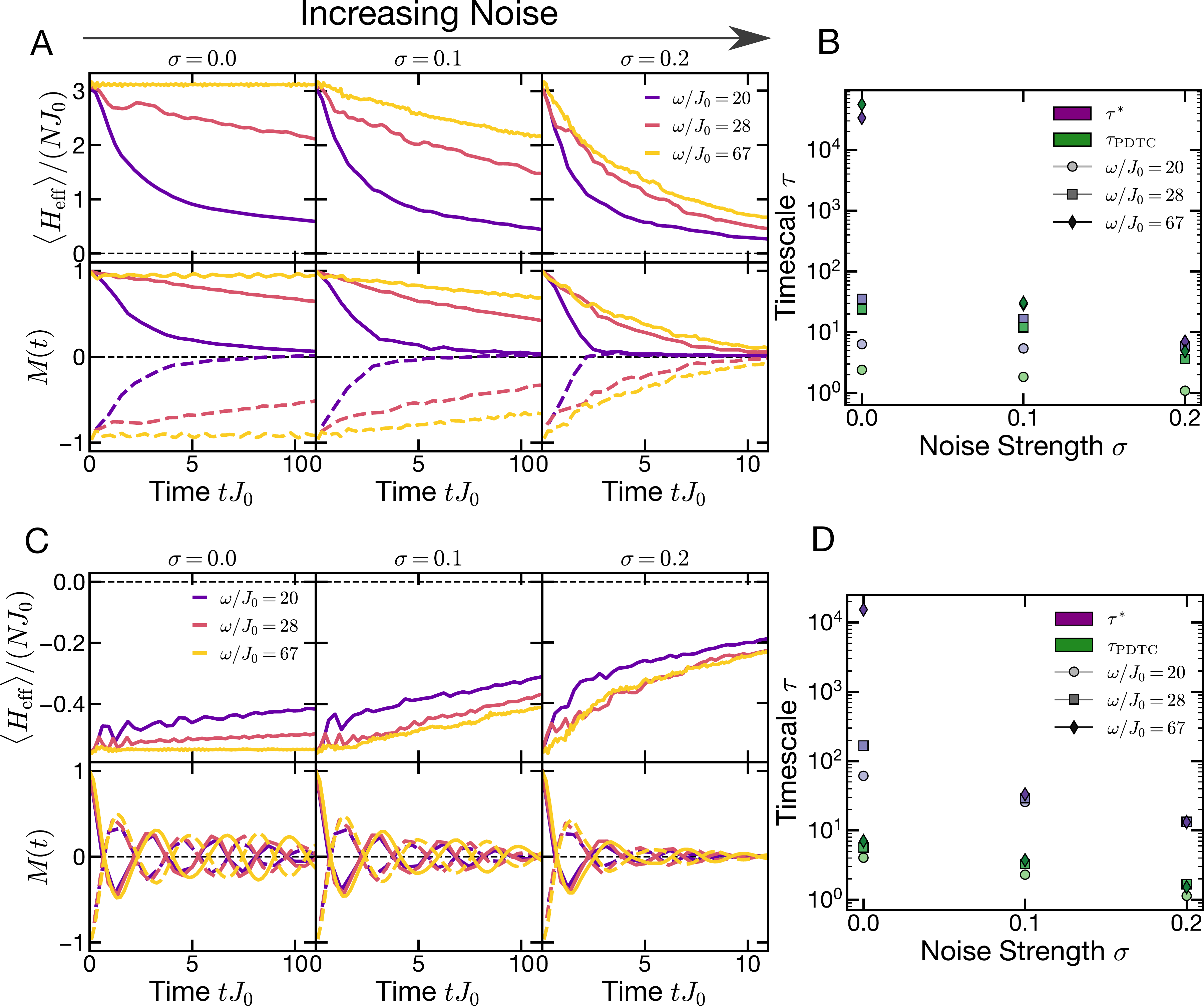}
    \caption{\small
    \textbf{Impact of noise on the PDTC and trivial Floquet dynamics}
    Heating dynamics of the polarized state (\textbf{A-B}) and the N\'{e}el state (\textbf{C-D}) for different strength of the noise $\sigma$. 
    The presence of noise hastens heating to the infinite temperature state, setting an upper bound on the time scales.
    Nevertheless, the trivial dynamics can be distinguished from the PDTC behavior.
    In the former, the magnetization decay remains frequency independent and much faster than the energy decay.
    In the latter, the magnetization decay has a similar frequency dependence to the energy density decay.
    }
     \label{fig:Noise}
\end{figure*}

\begin{figure*}
    \centering
    \includegraphics[width =4in]{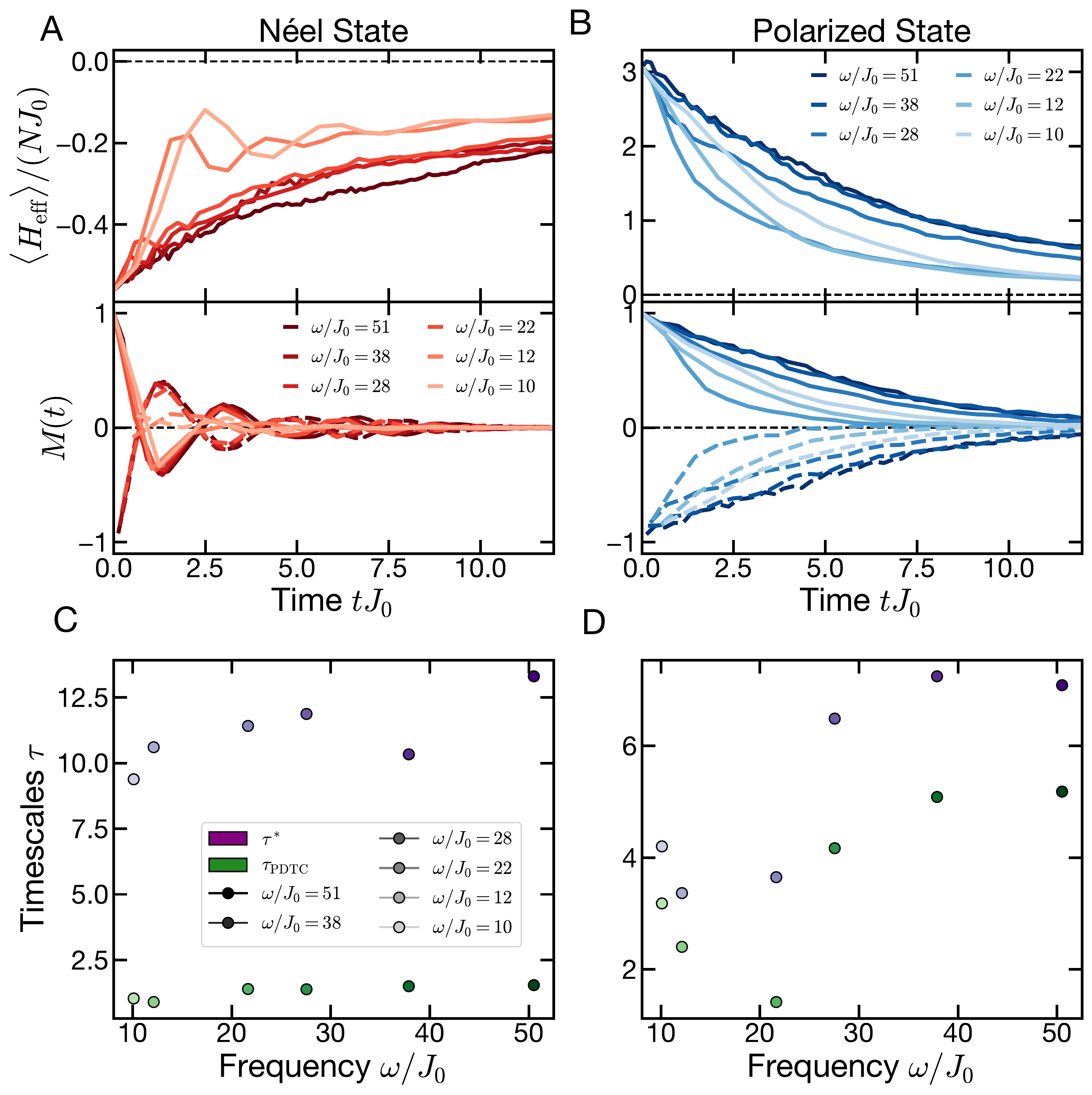}
    \caption{\small
    \textbf{Frequency dependence of the heating $\tau^*$ and time-crystalline $\tTC$ timescales for fixed strength of the noise.}
    \textbf{A-B.} Energy (top) and mangetization (bottom) dynamics for the N\'{e}el and the polarized initial state, in the presence of a fluctuating $B_z$ field with moderate strengths $\sigma = 0.1$.
    \textbf{C-D.}
    Upon increasing the frequency of the drive, we observe an increase in the heating time scale $\tau^*$ for both initial states, up until very high-frequencies where heating becomes dominated by the fluctuating noise.
    Crucially, the dynamics of the time-crystalline order is very different.
    For the N\'{e}el state, the time-crystalline order parameter decays quickly and is frequency independent---the system is in a trivial Floquet phase.
    For the polarized state, the increased heating time is mirrored by a increase of lifetime of the time-crystalline order parameter---the system is in the PDTC Floquet phase.
    }
    \label{fig:NoiseFreq}
\end{figure*}

We now turn to simulating the effect of noise in the observed dynamics.
As mentioned in Error sources, the most significant noise arises from laser power fluctuations.
We parameterize these fluctuations at the ions' location with the random variable $\epsilon(t)$, characterized by a flat spectrum and standard deviation $\sigma$ over a given duration (we have found that results are independent of the upper frequency cutoff as long as it is much larger than the drive frequency) .
We model this effect in the dynamics by adding a time dependence on the different parameters, $B_y$, $B_z$ and $J_{ij}$:
\begin{align}
B_y(t) &= B_y^{\textnormal{static}} \times [1 + \epsilon(t)] \\
B_z(t) &= B_z^{\textnormal{static}} + \epsilon(t) \times 2\pi \times 8 \mathrm{kHz} \\
J_{ij}(t) &= J_{ij}^{\textnormal{static}} \times [ 1 + 2\epsilon(t)]
\end{align}
The dependence of each term on $\epsilon(t)$ is determined by the
way that the laser power relates to that term, eg.~the $B_y(t)$ field depends linearly on the 2-photon Rabi frequency $\Omega$, which is proportional to laser power.
In Fig.~\ref{fig:Noise}, we highlight the effect of noise in two states at opposite sides of the spectrum and also considered in the main text: the polarized state and the N\'{e}el state.
The effect on either is qualitatively similar: noise reduces the frequency control of the Floquet heating leading to a plateau in the achievable $\tau^*$, in agreement with the experimental observation (we highlight this feature in Fig.~\ref{fig:NoiseFreq}). 
It is important to emphasize two points.
First, while for large frequency the noise dominates the heating towards the infinite temperature state, the frequency still provides a control of the Floquet heating highlighting the importance of of the drive.
Second, the dynamics of the magnetization is distinct in the trivial and the PDTC regimes.
In the former, the magnetization quickly decays before the heating time scale while in the latter, the magnetization decay follows the energy density decay.

\subsection*{Computing the crossover between trivial and PDTC behavior}

\begin{figure*}
    \centering
    \includegraphics[width = 0.7\textwidth]{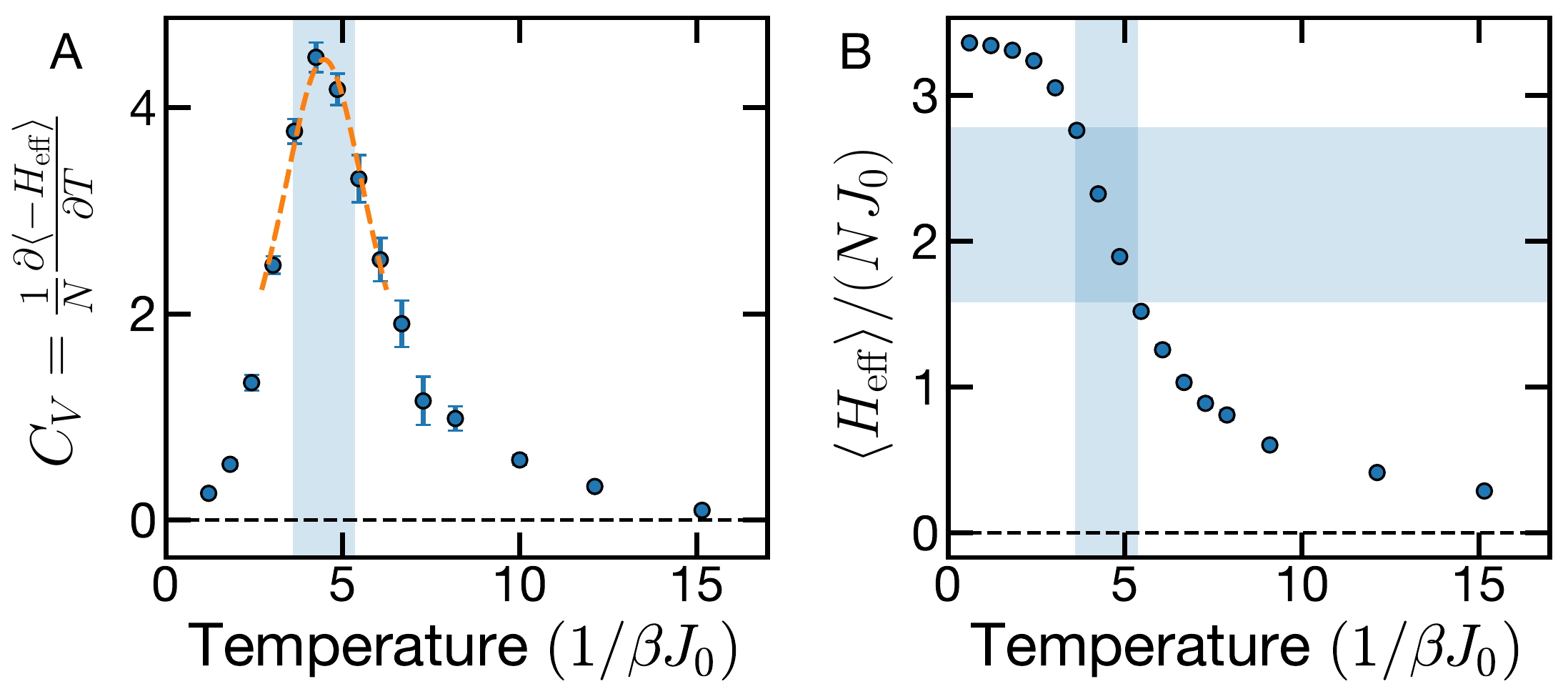}
    \caption{\small
    \textbf{Calculation of the para-to-ferromagnetic crossover region in $\Heff$.}
    Quantum Monte Carlo calculation for the $N=25$ spin chain considered in the experiment.
    \textbf{A.} We can locate the location and width of the crossover by characterizing the peak in the heat capacity $C_V$.
    \textbf{B.} Armed with the crossover temperature, we can directly map it into the crossover energy density which yields: $\Heff/(N J_0) = 2.18 \pm 0.60$.
    }
    \label{fig:QMC}
\end{figure*}

Since the PDTC behavior is dependent on a spontaneous symmetry-broken phase in $\Heff$, we can calculate the boundary between trivial and PDTC behavior by mapping the location of the transition (which for finite system size will emerge as a crossover).
In particular, owing to the antiferromagnetic interactions $J_{ij} >  0$, we are interested in the properties of the top of the spectrum, where the system orders ferromagnetically.
To this end, we perform a quantum Monte Carlo simulation of the Hamiltonian $-\Heff$ in an $N=25$ spin chain (Eq.~\ref{eq:H0}), which maps the calculation to the more common problem of finding the low-temperature phase boundary of the para-to-ferromagnetic phase.
In Fig.~\ref{fig:QMC} we present the results in terms of the original energy density.
By the system extending into the imaginary time dimension, we can perform the calculation at finite temperature $1/\beta$ and extract the crossover temperature and then map it to the relevant crossover energy density.
In particular, we identify the location of the crossover by the location of the peak in the heat capacity of the system which we fit to a Lorentzian peak, while the width is taken to be the quarter-width half-max of the peak.
The crossover location is then given by: $\varepsilon/J_0 = 2.18 \pm 0.60$.

\clearpage


\end{document}